# Defects in ZIF-8 crystallization and their impact on mechanical properties

*Annika F. Möslein [1], Lorenzo Donà [2], Bartolomeo Civalleri [2] and Jin-Chong Tan [1]\**

[1] Multifunctional Materials and Composites (MMC) Laboratory, Department of Engineering Science, University of Oxford, Parks Road, Oxford OX1 3PJ, U.K.

[2] Dipartimento di Chimica, Università di Torino, Via P. Giuria 5, 10125 Torino, Italy

ABSTRACT

 The growth process of metal-organic frameworks (MOFs) defines their properties for functional applications. However, it is very plausible that defects may occur during the crystallization of even seemingly perfect MOFs, such as ZIF-8, and yet, direct probing of such structural defects has been challenging due to the lack of techniques to locally examine individual nanocrystals. Now, we directly study local defects – such as missing linkers or metal vacancies – in ZIF-8 nano- and microcrystals with near-field infrared nanospectroscopy combined with density function theory calculations. We have tracked the chemical changes during crystallization and found that structural defects like zinc-rich regions gradually disappear with the ripening of the crystals, while missing linker defects prevail. The resulting open metal sites reduce the Young's modulus, as measured with tip force microscopy and supported by theoretical modelling, but also open the door for defect engineering to tune the adsorption and catalytic performance of ZIF-8.

INTRODUCTION

At the nanoscale, metal-organic frameworks (MOFs) crystals feature miscellaneous shapes and sizes. Their diversity originates in their vast physical and chemical properties, paving the way for applications in sensing technologies, drug delivery, gas capture, or catalysis, among others.[1-5] This multifunctionality emerges not simply because metal-organic frameworks are *per se* hybrid materials, built from metal clusters and organic linkers with a boundless number of possible combinations, but mostly because of their exceptional porosity, which sparks the adsorption, encapsulation or release of versatile guest molecules.[6, 7] The power of engineering MOFs for application is, therefore, ascribed solely to the meticulous control of the framework properties and its interactions, which essentially originates in the material synthesis.[7] For instance, the zeolitic imidazole framework ZIF-8, one of the most well-studied frameworks due to its stability and ease of synthesis, is obtained by combining zinc nitrate hexahydrate ($Zn(NO_3)_2 \cdot 6H_2O$) and 2-methylimidazole (mIm) ligands.[8] The material is then formed by self-assembly, a process where the basic building blocks – metal and linker – aggregate spontaneously to form a highly ordered 3-D structure: the crystalline framework. While the ZIF-8 crystals may materialize as nano- or microcrystals exhibiting rounded or faceted shapes, the development of size- and shape-controlling syntheses, in turn, benefits considerably from a detailed understanding of the crystallization process.

In particular, the formation of the ZIF-8 crystals, or MOF crystals in general, can be divided into the nucleation and growth process.[9] Since the nucleation is driven by random fluctuations in the bulk solution without dedicated nucleation sites, little can be done to tailor this mechanism. The crystal growth, on the contrary, can be manipulated by changing the molar ratio of reactants, amount of solvent, and other stimuli to yield crystals with desired sizes and shapes. Indeed, several studies scrutinized the impact of growth time, temperature, or the use of different modulators on the formation of ZIF-8 crystals by employing techniques such as atomic force microscopy (AFM) or scanning electron microscopy (SEM) combined with diffraction analysis or adsorption-desorption analyses to capture the size distribution, shape and crystallinity of the crystals. [10-15]

Yet, if seen in the context of controlling the material properties, another factor plays a crucial role in crystalline materials: the presence of defects.[16] Defect engineering opens pathways to locally tune the intrinsic porosity, create open-metal sites and modulate surface properties of MOFs, which has significant implications for separation, gas capture, catalysis, and mechanical responses.[17-24] Perhaps the model system for defective MOFs is UiO-66 (Universitetet i Oslo); here, nanoregions of ordered defects had been observed, which inspired detailed studies on the defects and their effects on the material.[19, 25-28]

Little is known, in contrast, about defects in ZIF-8; it may be partly because this material is considered to be amongst the most stable frameworks – but while this is true, it is in fact very plausible that defects or other irregularities may occur during the crystal growth, possibly with a significant impact on the performance of the materials, and its subsequent long-term stability.[29] So far, only the feasibility of local defects in ZIF-8, such as linker and metal vacancies, or dangling linkers, has been shown by computational modelling.[30, 31] However, despite a few other studies have focused on the implications of defects in gas separation[32] and storage[33], or found surface terminating defects[34], many open questions remain. One might ask whether defects can occur and transform during crystallization of ZIF-8? And to what extend do the defects affect the properties and as such, the performance of the material for potential applications?

To answer these questions, we employ a scattering-type scanning near-field optical microscope (s-SNOM), merging atomic force microscopy (AFM) with infrared spectroscopy to enable near-field infrared nanospectroscopy. It is this combination which can unravel the fine-scale features, mechanisms, and chemical interactions of MOFs at the nanoscale by yielding a Fourier Transform Infrared (FTIR) spectrum of a local 20 nm spot.[35-39] Previously, we have demonstrated the capability of this technique to probe individual MOF-type nanocrystals, such as ZIF-8, opening the door for discovering their fascinating characteristics from a new perspective: the single crystal level.[37] Compared with conventional techniques, this non-destructive approach allows not only imaging but simultaneously measuring the sample's properties including chemical information or physical properties at a resolution akin to AFM.[38] For instance, the mechanical properties can be obtained with the same setup, albeit operated in contact-mode instead of tapping mode. In these tip force microscopy (TFM) measurements, a force-distance curve is attained from every pixel of the AFM scan by retracting the AFM tip, measuring the required force to do so. The local stiffness is then determined from the force difference between the maximum force and the reference trigger point on the curve.[40] From the measured stiffness data of a sample surface, an image of the Young's modulus map can be derived. By simultaneously measuring the shape, size, chemical composition and mechanical properties at the nanoscale, the combination of these techniques – s-SNOM, nanoFTIR, AFM and TFM - allows us to shed new light on the growth process of the ZIF-8 nanocrystals.

In this study, we explore how defects transform during crystallization of ZIF-8, further elucidating their impact on the material performance. First, we study ZIF-8 nanocrystals before turning to microcrystals, obtained through two different synthesis routes (see Methods). To examine intermediate steps of the crystallization, the crystal growth was stopped after 1, 3 and 60 minutes by removing some material from the same batch, and performing three washing steps, respectively.

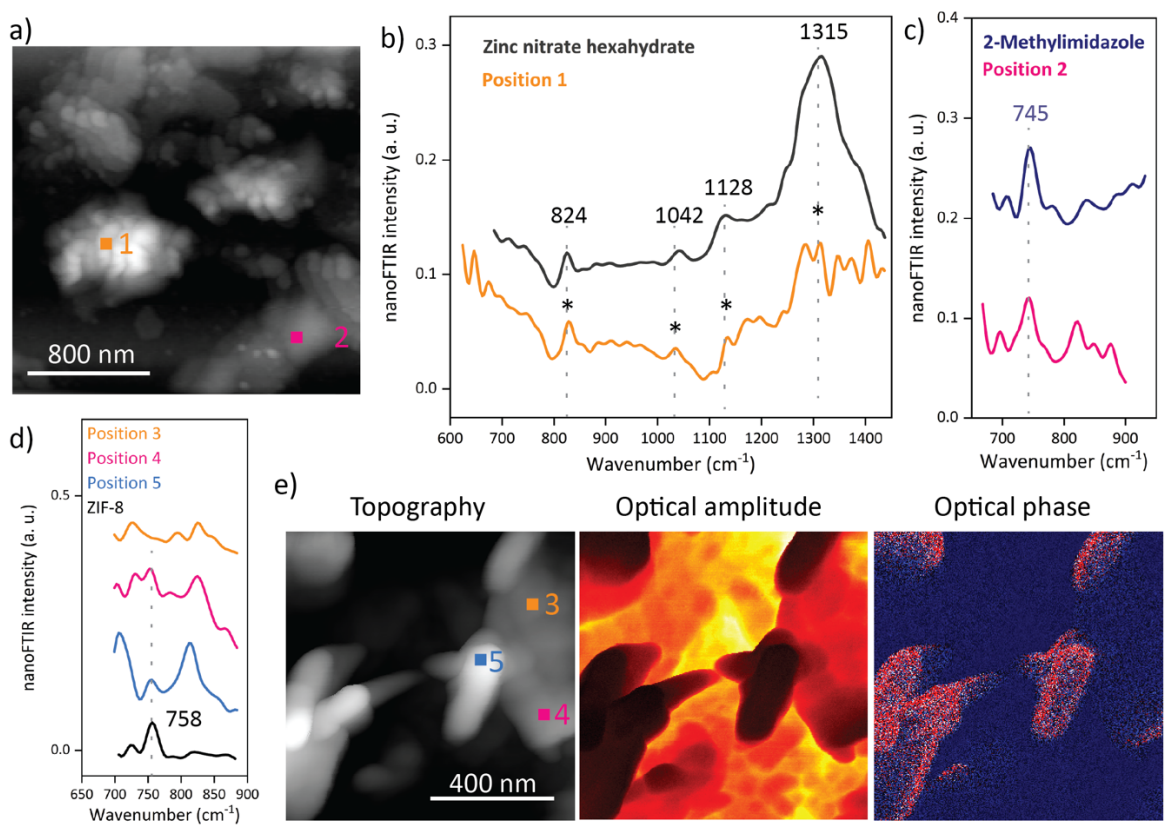

**Figure 1.** Near-field infrared spectroscopy of ZIF-8 with 1 minute growth time. a) The AFM image indicates the position, where the local nanoFTIR spectra (b,c) are measured, which resemble the spectra of the reactants. However, the emergence of a characteristic ZIF-8 peak (d) is already observable by locally probing at the positions shown in the AFM image (e), and local variations are further revealed in the optical amplitude and optical phase images (e).

RESULTS

**Inhomogeneity after a short growth time**

After 1 minute of growth time, the sample is dominated by inhomogeneous regions, although small, rounded shapes are already observable at the nanoscale (Fig. 1). Local probing of these nanoscopic morphologies reveals their spectral resemblances with the metal source, confirmed by the presence of the characteristic peaks of zinc nitrate hexahydrate at 824, 1042, 1128, and 1315 cm$^{-1}$ in the nanoFTIR spectrum (Fig. 1a,b). In contrast, the planate areas of the sample exhibit spectral features of the 2-methylimidazole (mIm) linkers, such as the distinctive vibrational band at 745 cm$^{-1}$ (Fig. 1a,c). Of course, it is the brief crystallization, and the lack of successfully formed nuclei and subsequent Ostwald ripening[41], that explain these distinct regions, where either the uncoordinated linker or the metal reactant dominates without evolved interactions between them. Only a closer examination reveals local variations as illustrated in the optical amplitude and phase images, which qualitatively contrast materials with different optical properties (Fig. 1d,e). Interestingly, the nanoFTIR spectra measured at different local spots with 20 nm resolution show contributions from both materials, even displaying the emergence of the characteristic peak of ZIF-8 at 758 cm$^{-1}$, assigned to an out-of-plane ring mode of the framework (Fig. 1d). This finding confirms not merely the growing chemical interactions and bond formations, but the formation of the framework itself, which, at least partially, is beginning to crystallize within the first minute. Nonetheless, the large amount of uncoordinated linker and metal salt may indicate which defects could materialize during further crystallization.

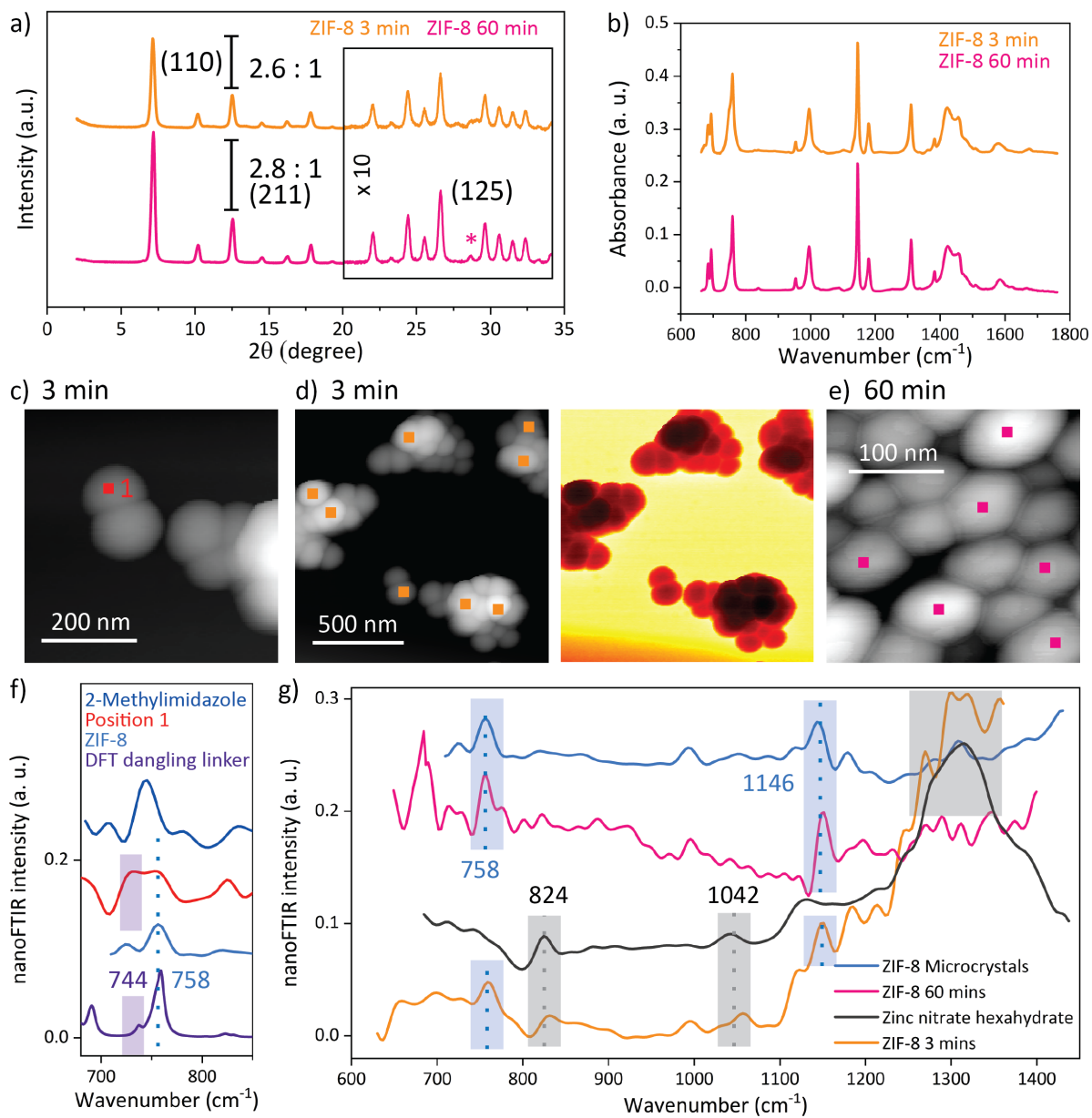

**Figure 2.** Comparison between ZIF-8 nanocrystals obtained after 3 and 60 minutes. a) The XRD pattern confirms the crystallinity of the sample obtained after 3 minutes. b) ATR-FTIR spectra reveal that ZIF-8 crystals are formed after 3 minutes. c,d,e) Imaging the topography and optical amplitude of nanocrystals after 3 minutes growth time with s-SNOM (c,d) resembles the AFM image of the final growth stage of the ZIF-8 crystals (e). f) nanoFTIR spectra of a 20 nm spot, as indicated in the AFM image (c), compared with the spectra of linker, ZIF-8, and simulated FTIR

spectrum of ZIF-8 with dangling linker defect. g) The average of several local spectra (positions illustrated in d,e) reveal local variations, which vanish with prolonged crystallization.

**Defects gradually disappearing with prolonged crystallization**

After a growth time of only 3 minutes, the ZIF-8 nanocrystals have already formed, confirmed by comparing their X-ray diffraction (XRD) pattern and attenuated total reflection (ATR) FTIR spectra with the ones measured on crystals after a growth time of 60 minutes (Fig. 2a,b). In particular, the XRD Bragg peaks are fully resolved, matching the reported characteristic peaks at $2\theta$ values of 7.3, 10.3, 12.6, 16.4, and 17 degrees, and thus clearly revealing crystallinity for the nanocrystals attained after 3 minutes (Fig. S1).[8] Here, the high intensity of the (110) peak at 7.3º is attributed to the formation of ZIF-8 with a regular rhombic dodecahedron shape, which resembles the final stage of the growth of ZIF-8 crystals.[14] Small variations, however, can be detected, such as the lacking of the (125) peak at 28º, or a changing relative intensity of the (110) : (211) planes, corresponding to the two most intense diffraction peaks. Albeit minimal, a rise in the intensity of the (110) plane accompanied by a narrowing of the full width at half maximum (FWHM) from 0.299 to 0.288 is observed with prolonged crystallization, indicating a stronger long-range ordering, and thus, implying a higher crystallinity (Fig. S2). At first glance, these findings are in good agreement with the previously reported result, where the crystallinity increased with the synthesis duration due to the Ostwald ripening process.[13-15] It has been suggested that 5 minutes was insufficient to grow ZIF-8 crystals[13], but here, paradoxically, it is the marginal increase of crystallinity, and the ATR-FTIR spectrum of the crystals with 3 minutes growth time, which matches the one of the final growth stage, that indicate the self-assembly of ZIF-8 nanocrystals after only 3 minutes. [13-15]

Though these bulk measurements suggest the growth of ZIF-8 nanocrystals after a relatively short crystallization time, near-field infrared spectroscopy reveals that, in fact, these crystals are far from defect-free. For instance, the average of several local nanoFTIR spectra measured on individual crystals exposes the strong contribution of zinc nitrate hexahydrate with its characteristic peaks at 824 and 1042 cm$^{-1}$ (Fig. 2d,g). As opposed to ATR-FTIR, where the spectra are obtained from an average of the bulk, polycrystalline material, the nanoFTIR spectrum is probed locally with a probing depth of only a few nanometers, or tens of unit cells. Therefore, the presence of vibrational modes associated with the metal reactant truly indicates local variations close to the surface of the framework, such as different termination groups or under-coordinated metal clusters; in other words: Zn-rich regions. One reason for the predominant appearance of the metal modes in the vibrational spectrum averaged over several positions could stem from the recurring defect of missing linkers, but the existence of additional metal clusters seems to be more likely, especially since previous studies have shown that Zn-rich regions can emerge on the outermost surface of ZIF-8.[34, 42] Creating open metal sites through defects can be leveraged as a technique to enhance the reactivity of framework materials, be it in catalytic applications or gas adsorption. Here, the under-coordinated Zn ions close to the crystal surface present additional reaction sites; precisely this feature, however, may alleviate the long-term stability of the material by deteriorating its hydrophobicity.

Turning even more locally, individual point spectra show the superposition of peaks assigned to ZIF-8 (758 cm$^{-1}$) and the mIm linkers (745 cm$^{-1}$), further uncovering local defects associated with the linker, such as partially coordinated linker, or dangling linker, as the termination units close to the crystal surface (Fig. 2c,f). This is confirmed by density functional theory (DFT) calculations using the CRYSTAL code.[43] A defective ZIF-8 crystal was modelled, where a metal vacancy introduced dangling linker groups (see Methods). In the simulated FTIR spectrum, an additional peak appears at 744 cm$^{-1}$ which is assigned to the vibrational modes of the dangling linker. Of course, such irregularities close to the external surface of the crystallites are to be expected when looking at materials on the nanoscale. However, none of these structural defects are observed in the nanoFTIR spectrum of the nanocrystals with a growth time of 60 minutes. Instead, the local measurements are resembling the average spectrum, and neither reveal local variations, nor individual contributions of the reactants (Fig. 2g, SI Fig. S3). This leads to the conclusion that structural defects, such as the different terminations, gradually disappear with prolonged crystallization.

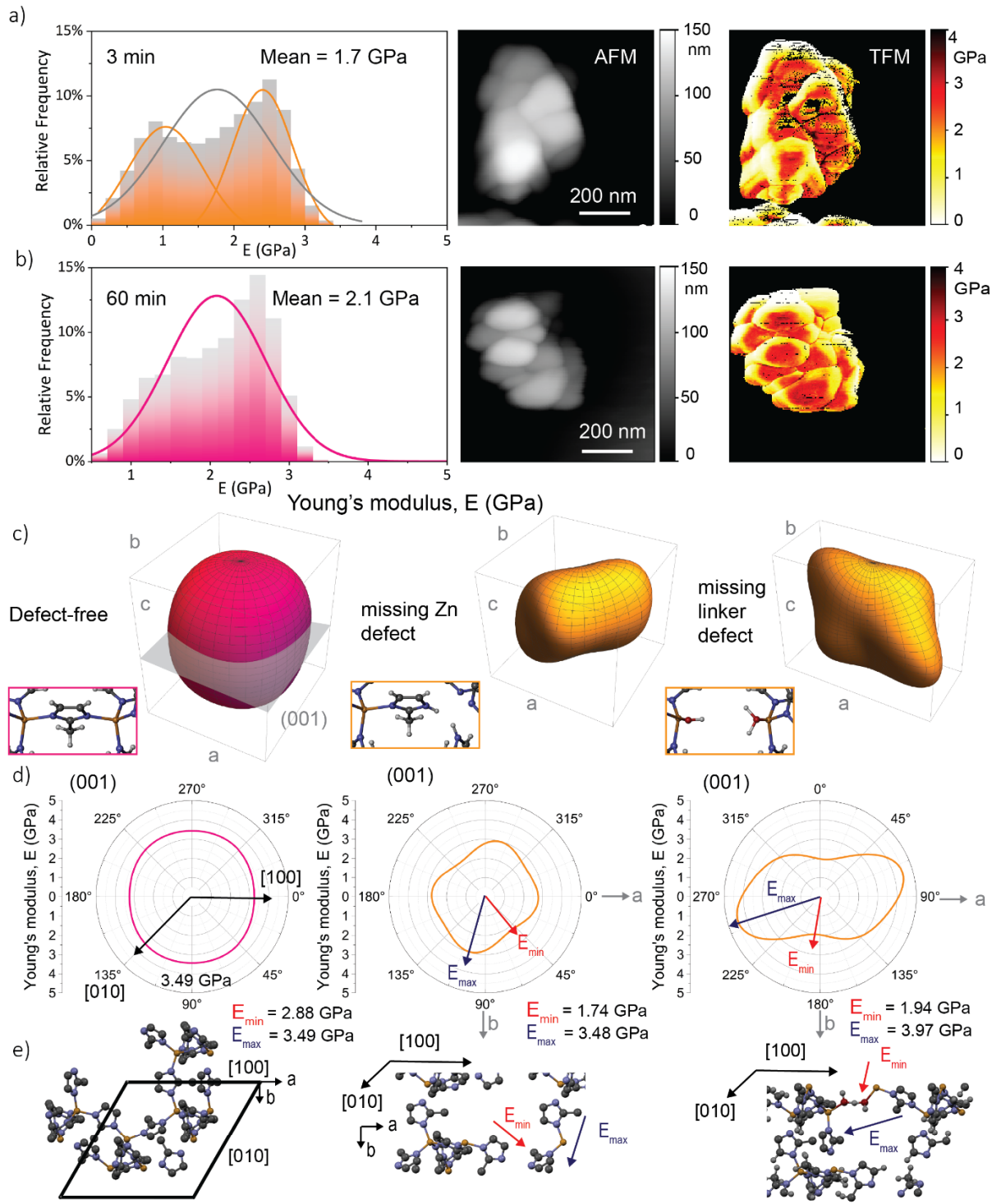

**Figure 3.** Tip force microscopy (TFM) on ZIF-8 nanocrystals with different growth time (top: 3 minutes, bottom: 60 minutes). a,b) Histogram and normal distribution of Young's modulus corresponding to data collected on the nanocrystals on each pixel, along with AFM images, and mapping of the Young's modulus, measured with tip force. c) Young's modulus representation surface in 3-D spherical coordinates, along with schematic of ZIF-8 and defective structures. d) Polar plots projected onto the (001) planes. e) Schematics of the structure-property relationships indicating the orientations of maximum and minimum Young's moduli. Color code: dark grey: carbon, light grey: hydrogen, blue: nitrogen, red: oxygen, orange: zinc.

**Mechanical properties change with vanishing defects**

Of course, the different termination units, the under-coordinated metal, and the dangling linker – all of these invite the question of how and to what extend they might influence the properties of the nanocrystals, which are so intrinsically linked with the framework assemblage, arrangements of functional groups, and characteristics of the pore. Even if only slightly, defects may affect the material's stability by disrupting the crystalline order of the framework material, but whether the relation between defects and their impact on the material performance can be precisely established remains to be studied. This is best done with tip force microscopy (TFM), where the Young's modulus $E$ (i.e. ratio of uniaxial stress over strain in elastic regime) is measured at every pixel of the scan, while simultaneously imaging the topography at nanoscale resolution. Hereby, the local stiffness of the sample under investigation is acquired with a resolution akin to AFM imaging (Fig. 3).[40] It is worth mentioning that the artifacts in the measurements (black lines) have been filtered out, before deriving the mean stiffness of the nanocrystals (Fig. 3a,b). Though the value of the stiffness is generally lower than previously observed for ZIF-8 crystals, be it with indentation or AFM nanoindentation, at such small scales – as measured with TFM – it reflects the local nature of individual nanocrystals.[44] Thus, given the same calibration, these TFM measurements can be employed for comparison between different materials. For the ZIF-8 crystals with a growth time of 3 minutes, the Young's modulus is attained with a mean at 1.7 GPa (standard deviation of 0.7

GPa), which is significantly lower when compared to the stiffness of the crystals at the final growth stage with the mean at 2.1 GPa (standard deviation of 0.5 GPa). Likewise, the crystals with short crystallization unveil more local variance in the Young's modulus, even revealing a bimodal distribution (Fig. 3b, Fig. S4). A simple way of putting this is to say that higher crystallinity and thus stiffness is achieved with the prolonged crystallization, but there is more to it than that, especially since the increase in crystallinity is marginal, as discussed earlier. An alternative explanation might be to suggest that the observed structural defects, be they different termination units, uncoordinated metal clusters or dangling linkers, introduce local disturbance to the otherwise periodic bulk structure and thus disrupt the stability of the framework as a whole; in other words, one might describe them as local disorder with far-reaching impact. This applies particularly to the mechanical properties due to their dependencies upon the long-range order of the crystal. The observed phenomenon that stiffness, or mechanical stability in general, is lowered with increasing defect level, coincides with previously reported simulations, where the elastic constants were calculated for defective UiO-66 crystals.[45]

This tendency is further confirmed by our DFT calculations with the CRYSTAL17 code to compute the elastic tensors for three different structures of ZIF-8: first, the ideal, defect-free crystal was simulated using the PBEsol0-3c method.[43,46] Derived from this ideal unit cell, the missing zinc defect was introduced as a zinc vacancy, and replacing two N-Zn bonds with N-H bonds, thereby creating dangling, or undercoordinated linkers. The third system that, in turn, simulates the missing linker defect, was attained by removing a linker group, while the two unsaturated metal sites were filled by an associating water and the conjugate base of the proton-donating group. By visualizing the elastic tensors, and the associated mechanical properties derived from these (SI Tables S1, S2, and Figs. S5-S8), the experimental findings can be explained. As shown in Fig. 3 c and d, the Young's modulus of the defect-free crystal is highly isotropic, consistent with a previous study on the elastic constants of ZIF-8 single crystal.[47] On the contrary, with the introduction of a missing zinc defect, an anisotropic behavior is revealed. Although the maximum stiffness – in direction of the stable, undisturbed zinc bonds – is akin to the one of the perfect crystal, the minimum Young's modulus is significantly smaller (Fig. 3e). If forces are applied in the direction of the longest pore width, the dangling linker groups can now rotate or twist due to the missing zinc defect, which decreases the framework stiffness accordingly. Similarly, the effect of increased anisotropy due to defects can be noticed in the model of the missing linker; in fact, it is this system where it becomes most evident. Here, a bond between the associated water and conjugate base is formed to yield a mechanically stable system. This, however, changes the conformation of the pore significantly: in direction of this additional bond, the framework stiffness is increased due to shorter bond lengths and associated pore conformation. As a consequence, the maximum Young's modulus is even larger than in the defect-free case. Yet, in the vertical direction, these weak hydrogen bonds are easily disrupted under compression or extension forces, leading to a much larger pore than in the defect-free crystal, which describes the decrease in the minimum Young's modulus.

The theoretical results above can explain the observed characteristics of the local stiffness measured with TFM. Here, it is worth mentioning that the calculations assume a periodic occurrence of the introduced defects, while, in reality, the number of defects and their orientation is – most likely – random. Hence, in the crystals with more defects, the increased anisotropy is the reason for the larger variance observed in mechanical properties, or even bimodal distribution of the Young's modulus, while the average stiffness, and as such, the material's stability is generally reduced in the defective structures. If inverted, the effect of this is that structural flexibility, adsorption, or other responses to external stimuli can be enhanced in an otherwise 'rigid' MOFs (greater stiffness) by introducing structural defects.

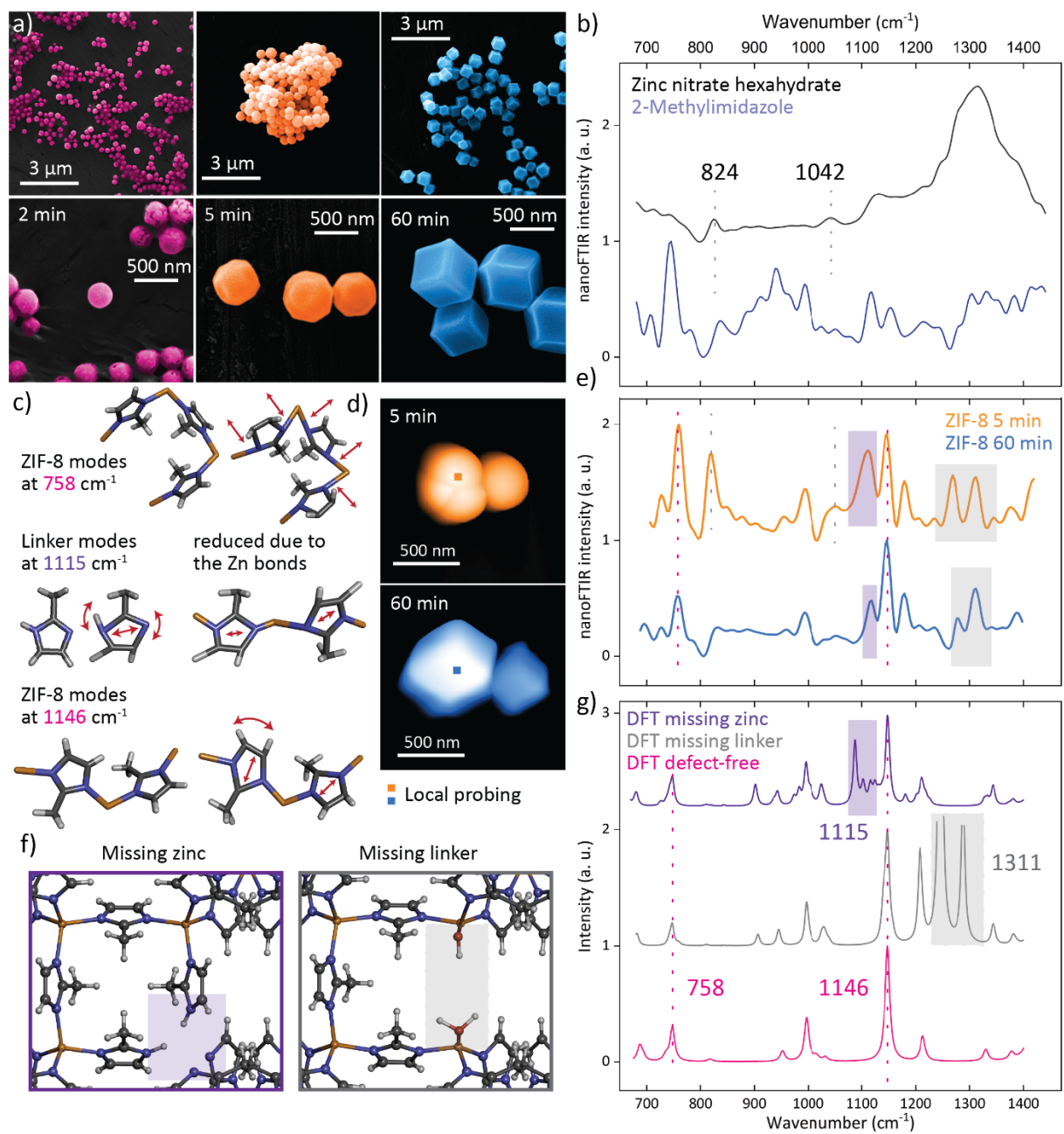

**Figure 4.** Transformation of ZIF-8 microcrystals. a) SEM imaging of ZIF-8 crystals with different growth times. b) nanoFTIR spectra of the reactants. c) Schematic representation of key vibrational

modes. d) AFM images of ZIF-8 crystals obtained after 5 and 60 min indicating positions of local probing. e) Corresponding local nanoFTIR spectra. f) Schematic representation of the simulated defects. g) Corresponding simulated FTIR spectra compared with a perfect ZIF-8 crystal.

**Defect Transformation in Microcrystals**

Similar observations are made by tracking the changes in key vibrational bands during the crystallization of ZIF-8 microcrystals, which further reveals how defects transform. As shown in the SEM images in Figure 4 a, the crystal shape evidently changes from the round morphology obtained after 2-minute growth time, towards exhibiting the first facets after 5 minutes, and eventually reaching the rhombic dodecahedron shape at the final growth stage after 60 minutes. This is accompanied by a gradual increase in the averaged crystal size from 200 to 500 nm. A comparison of the two nanoFTIR spectra for the 5 minutes and the 60 minutes growth time with the DFT calculation depicts a close resemblance of the samples yet unravelling several salient changes in peak positions and intensities (Fig. 4a). First, the nanoFTIR spectrum corresponding to the longer growth process shows a better match with the calculated spectrum of an idealized, periodic crystal, since, with the prolonged crystallization, the number of ideal bonds increases, and thus the positions and intensities resemble the calculated IR spectrum. Second, a detailed study of the individual vibrations, considering the DFT calculations as well as the nanoFTIR spectra of the reactants before synthesis, gives insights into both the structural and chemical changes happening during the crystallization of the 3-D framework. The pronounced peak at 758 cm$^{-1}$, for instance, correlates with the symmetric out-of-plane bending of the mIm ring and associated motions of the H-C=C-H bond present in the framework structure (Fig. 4c). Conversely, the peak at 824 cm$^{-1}$ is related to the vibrations of the metal salt (zinc nitrate hexahydrate), which is not expected to appear after the construction of ZIF-8, where both reactants are used up entirely to form ZIF-8 crystals.

However, since the associated vibrational mode is still present in the nanoFTIR spectrum, the crystals formed after only 5 minutes still contain residuals of the excess reactants, thus suggesting that some metal clusters are not fully coordinated with the mIm linkers. Given that extensive washing steps have been performed for all samples, one can conclude the metal cations are, albeit poorly coordinated, bound to the framework. There is some doubt where precisely the metal clusters could be located, but the strong appearance of the vibrational mode in fact indicates a repeating pattern of this defect close to the crystal surface; most likely, it is the defect-terminating zinc clusters with missing linkers, which could further explain why the crystals, rather than resembling the stable, rhombic dodecahedron shape, only slightly imply facets after 5 minutes growth time. In comparison with the characteristic peak of ZIF-8 at 758 $cm^{-1}$, the vibrational mode associated with the metal reactant at 824 $cm^{-1}$ depicts a reduced relative intensity from 0.74:1 to 0.48:1 with a longer time of crystallization; a finding which already contains, in essence, everything that has been observed for the nanocrystals. With prolonged growth time, the defects of additional metal clusters gradually disappear. The same phenomenon is revealed in the reduction of the less pronounced peak at 1042 $cm^{-1}$; likewise, this peak simultaneously vanishes as the complete framework is assembled (Fig. 4 b,e).

So far, the emphasis fell on the defects attributed to the additional metal cluster, however, it is further evident that the defect of dangling linker can be detected at such local scales. In particular, the vibrational mode at 1115 $cm^{-1}$, a peak associated with asymmetric in-plane ring stretching of the uncoordinated linker C-N-H bonds, is strongly present in the early stage of the crystal growth process. Interestingly, these modes appear in the simulated spectrum of a defective ZIF-8 crystal, where dangling linkers have been introduced by virtue of a metal vacancy. Meanwhile, in the final stage of the crystal growth, this mode has reduced in its intensity, since the N-H bonds in the linker are replaced by the stable N-Zn bonds suppressing this vibration (Fig. 4c).

Thus, in the 5-min crystallization, the high relative intensity of this vibration with 0.85 : 1 when compared with the ZIF-8 peak at 1146 cm$^{-1}$ – the latter attributed to the C-H rocking and ring stretching in the framework – reveals the defect of unsaturated, ill-terminating ligands, or so-called dangling linkers, which are only bound to one zinc atom. As a result, the free-space vibrations of the linker, particularly the mode associated with the unwieldy asymmetric ring stretching and the N-H bonds, are enhanced and thus detectable in the nanoFTIR spectrum. Once bound to two zinc atoms in a fully assembled framework, these motions are mostly constrained, as indicated by the disappearance of this mode, or the decrease in relative intensity to 0.47 : 1, with prolonged crystallization. In fact, the DFT assessment shows that this asymmetric ring stretching mode is even more suppressed in a defect-free crystal, as implied by the low relative intensity between the two simulated peaks (0.05 : 1). This suggests that, to some extent, and unlike the defects associated with the Zn-rich regions, this type of defect is found to prevail even when the final stage of the microcrystal growth is reached after 60 minutes. One can thus conclude that ordered dangling-linker defects can exist in ZIF-8 microcrystals, including those deemed seemingly perfect.

Thus there emerged the potential of measuring defects in ZIF-8, which is not by any means confined to the initial stages of early crystallization. The significance of this finding for the understanding of the prototypical, 'stable' ZIF-8 crystal is so great that it is worth discussing a little further. Whereas the bulk ATR-FTIR measurements perfectly match the calculated FTIR spectrum (Fig. S5), the local nanoFTIR spectra do not. Instead, it is possible to pinpoint local characteristics in the vibrational modes, and, in combination with computational modelling, associate them with chemical and structural peculiarities at the nanoscale. A glance at the vibrations between 1250 and 1350 cm$^{-1}$ will further illustrate this, as these additional vibrations are observed in the local spectra of ZIF-8 crystals and can be assigned to missing linker defects, yet, they neither appear in the ATR-FTIR measurement nor in the calculated crystal which is defect-free (Fig. 4e,g). But this is precisely what the nanocrystals are not, for they feature these modes, namely the missing linker defects, at such local scales. Akin to the DFT calculation, where the resulting two unsaturated metal sites are filled by an associating water and the conjugate base of the proton-donating group, the associated vibrational modes appear in the nanoFTIR spectrum, indicating the presence of such aqueous linker vacancy at the crystal surface. These results highlight that the defect sites may alleviate the hydrophobicity, and as such the long-term stability of ZIF-8, by exposing open metal sites to solvents, gases or other reactants. On the other side, the same phenomenon could also enhance the adsorption capabilities of the material - what had previously been regarded as structural defects or at best imperfections may come to be considered as merit.

DISCUSSION

In this work, we tracked the transformation of defects in ZIF-8 by locally probing nano- and microcrystals at different stages during the crystal growth with near-field infrared spectroscopy (s-SNOM). As opposed to established techniques which measure a spatial average over the bulk material, the use of s-SNOM yields chemical information with a resolution akin to AFM imaging. That way, the coexistence of defects can be observed during crystallization by pinpointing the vibrational dynamics of a 20 nm spot, or several unit cells. Whereas, after 1 minute growth time, large amounts of uncoordinated linker and metal reactants are still dominant, it is already possible to note the presence of round nanocrystals; here, local probing with nanoFTIR confirms the emergence of characteristic peaks of ZIF-8, thus suggesting that the framework is already forming after such a short crystallization time. Indeed, ZIF-8 nanocrystals which match the characteristics of the final growth stage are attained after only 3 minutes, as corroborated with conventional characterization techniques like XRD, ATR-FTIR and AFM. However, there is one distinction which can only be revealed with nanospectroscopy: that between ZIF-8 crystals with perfect periodicity and those with structural defects. For instance, the local nanoFTIR spectra show the dominance of Zn ions, or under-coordinated metal clusters, close to the surface of the nanocrystals with 3 minutes growth time, and even reveal the coexistence of vibrational modes associated with ZIF-8 and the uncoordinated linker at the same 20 nm spot, thereby indicating defect-terminating ligands. With prolonged crystallization, or the ripening of the crystals, these defects vanish, and the final growth stage is reached after 60 minutes. Perhaps the most significant implication of this defect evolution – it is even more striking than observing the defect itself – consists in the change of mechanical properties. A glance at the local Young's modulus, as measured with tip force microscopy (TFM), has illustrated this: not only is the stiffness significantly lower in the presence of defects, but it also exhibits a higher local anisotropy. These experimental findings are confirmed by theoretical DFT calculations, where the mechanical properties of ZIF-8 and its defective structures have been computed. These trends are attributed to the fact that defects introduce local disorder to the otherwise highly ordered 3-D framework, ultimately compromising the material stability, although their impact on the material properties might extend even further.

The same phenomenon of defect transformation during crystallization was observed in ZIF-8 microcrystals. While the individual crystals transform from a spherical morphology to the rhombic dodecahedron shape, the underpinning chemical changes were tracked on the single crystal level for the first time. After a growth time of 5 minutes, the faceted shape begins to emerge, yet several local structural defects are determined with nanospectroscopy; those include under-coordinated metal clusters and dangling linkers. Again, the trend of defect evolution is epitomized in their gradual disappearance with prolonged crystal growth time, although the defect of dangling linkers, if only slightly, is deemed to prevail. Similarly, defect sites of missing linkers lead to unsaturated metal sites, which adsorb water molecules, as confirmed with nanospectroscopy and DFT calculations. Out of that understanding grows the doubt whether the stable, faceted ZIF-8 crystals, which have typically been assumed to be essentially defect-free, are in fact perfect or whether the defects were just invisible to most of the characterization techniques. The combination of s-SNOM with DFT sparks the exploration of such local defects, not only in MOFs, but also in other crystalline materials. In addition, the use of TFM enables the link between the physical, chemical, and mechanical properties to shed new lights on the implications of defects, or in fact any features at the nanoscale. We envisage that these findings and techniques invite future studies on defects in MOFs and cognate framework materials, to either evaluate the stability of 'stable' MOFs for targeted application, or on the contrary, to leverage local defect-engineering for enhanced material performance.

METHODS

**Synthesis of ZIF-8 nanocrystals.** ZIF-8 nanocrystals were synthesized by dissolving 4.5 mmol of zinc nitrate hexahydrate ($Zn(NO_3)_2 \cdot 6H_2O$, 98%, SigmaAldrich) and 13.5 mmol 2-methylimidazole (mIm) (98%, Sigma-Aldrich) in 60 mL of methanol, respectively. After combining the two clear solutions, the white colloidal solution was rigorously stirred for 1 min and then left to form the nanocrystals. Immediately, some material was removed, diluted in fresh methanol and washed three times. Each washing step encompassed centrifugation at 8000 rpm for 5 minutes followed by solvent exchange with fresh methanol, and sonication for 30 seconds. The same procedure was repeated after 3, 5, and 60 minutes: after these time intervals, some material was removed and thoroughly wash to stop the growth process.

**Synthesis of ZIF-8 microcrystals**. Two precursor solutions were prepared by dissolving 4 mmol of zinc nitrate hexahydrate and 4 mmol of 2-methylimidazole in 40 mL of methanol, respectively. A mixture was obtained by combining the two precursor solutions, which was stirred for 1 minute and then left without stirring. After specific time intervals (2, 5, and 60 min), some material was removed from the batch and immediately washed three times with methanol and centrifugation.

**Sample preparation for nano-scale analytics.** Each sample was diluted in methanol and drop casted onto a silicon substrate. To eliminate any excess solvent, the sample was dried in a vacuum oven at 80 °C for at least 30 min. The spectra of the reactants were measured by dissolving zinc nitrate hexahydrate ($Zn(NO_3)_2 \cdot 6H_2O$, 98%, SigmaAldrich) or 2-methylimidazole (98%, Sigma-Aldrich) in methanol, respectively. Likewise, the solution was drop casted onto a clean silicon substrate, and dried at 80 °C for 30 min.

**Powder X-Ray Diffraction (PXRD).** PXRD patterns were measured at a step size of 0.02° and step speed of 0.01°/min using the Rigaku MiniFlex diffractometer equipped with a Cu Kα source, and validated against the simulated XRD pattern (CSD database code: VELVOY).

**ATR-FTIR.** Attenuated total reflection (ATR)-FTIR measurements on bulk material were performed using the ThermoFisher Scientific Nicolet iS10 FTIR spectrometer with a spectral resolution of 4 $cm^{-1}$.

**SEM imaging.** Scanning electron microscope (SEM) images of the samples were obtained with a TESCAN LYRA3 electron microscope. Backscattered electron and secondary electron SEM images were obtained at 10 keV under high vacuum. The false-color images were produced using Adobe Photoshop.

**nanoFTIR.** The near-field optical measurements were performed with a neaSNOM instrument (neaspec GmbH) based on a tapping-mode AFM, where the platinum-coated tip (NanoAndMore GmbH, cantilever resonance frequency 250 kHz, nominal tip radius ~20 nm) was illuminated by a broadband femtosecond laser. The coherent mid-infrared light was generated through the nonlinear difference-frequency combination of two beams from fiber lasers (TOPTICA Photonics Inc.) in a GaSe crystal. Laser A was selected for the measurements covering the range from 700 to 1400 $cm^{-1}$. Demodulation of the optical signal at higher harmonics of the tip resonance frequency eliminated background contributions to yield the near-field signal, comprising amplitude and phase of the scattered wave from the tip. Employing a pseudo-heterodyne interferometric detection module, the complex optical response of the material is measured, where the real part refers to the nanoFTIR reflectance and the imaginary part depicts the nanoFTIR absorption spectrum. Each spectrum was acquired from an average of 14 Fourier-processed interferograms with 10 $cm^{-1}$ spectral resolution, 2048 points per interferogram, and 14-ms integration time per pixel. The sample spectrum was normalized to a reference spectrum measured on the silicon substrate. All measurements were carried out under ambient conditions.

**Tip force microscopy (TFM).** Tip force microscopy is employed as an additional module to the neaSNOM instrument, but here the AFM is operated in contact mode. The z-piezo driver is modulated at a sinusoidal motion with an amplitude of 40 mV and modulation frequency of 610 Hz. A complete force-distance cycle is performed at this rate for each pixel with an integration time of 33 ms (200 × 200 pixels per image). The technique follows the descriptions of pulsed force mode.[40] Each cycle comprises an approach of the AFM tip from free oscillation until establishing contact with the sample, followed by the subsequent retraction. More specifically, the contact is established due to the (negative) attractive force between the tip and the sample surface. Once in contact, the piezo drives the tip even closer to the sample until the (positive) repulsive force reaches a maximum. Upon retraction of the tip, the repulsive force decreases and is replaced by the attractive force due to the adhesion between the sample and the tip until the contact is lost, and the tip freely oscillates. From this cycle, various properties can be derived.[40, 48, 49] For instance, the topography image is obtained from the maximum force, which is fed back to the control circuit to maintain a constant normal force. An adhesion image can be created based on the maximum adhesion force for each pixel. The local stiffness is attained from the force difference between the maximum force and a set point in the repulsive part of the force signal. Hard surfaces lead to a larger force difference than observed for soft surfaces. Calibration measurements were carried out on the silicon substrate, and a PS/PMMA polymer blend sample with known Young's moduli was used to validate the calibration. The mean local stiffness was obtained from the stiffness images by filtering out outliers and the background region containing the substrate. The normal distribution, as well as the bimodal distributions, were then calculated with the integrated analysis tools in OriginPro 2019.

**Density functional theory (DFT) calculations.** The theoretical vibrational spectra of ZIF-8 and defective ZIF-8 models as well as elastic constants calculation were computed with the PBEsol0-3c composite method.[46, 50, 51] Recently developed, PBEsol0-3c is a cost-effective composite method based on HF/DFT Hamiltonian combined with a double-zeta basis set, providing a good trade-off between cost and accuracy.[52] The calculations were carried out with the periodic *ab initio* CRYSTAL17 code running in MPP mode on ARCUS-B, part of the high performance computing facility at the University of Oxford, UK, and on the UK national HPC facility ARCHER2.[43] The missing metal (or so-called dangling linker) defect was created by removing a metal atom, and replacing two N-Zn bonds with N-H bonds similar to the uncoordinated linker. Removing a linker group, in turn, led to the missing linker defect, where the two unsaturated metal sites were filled by an associating water and the conjugate base of the proton-donating group. After geometry optimization, the Berry Phase approach was employed to calculate the IR intensities. Subsequently, a continuous spectrum was obtained by fitting the calculated IR intensities with Lorentzian peak shapes with a FWHM of 10 cm$^{-1}$.[53]

The calculated IR spectra were shifted using multiple scale factors, since this significantly improved the match with experimental data compared with the use of an overall scaling constant.[54] For the range from 600 to 850 cm$^{-1}$, the IR spectrum was scaled with a factor of 0.936, while the region between 850 and 1050 cm$^{-1}$ was scaled by 0.964. Higher wavenumbers were scaled by a factor of 0.958.

The single-crystal elastic constants of the elasticity matrix (tensor) were calculated using the numerical first derivative of the analytic cell gradients.[55] These values correspond to the independent elastic stiffness coefficients, $C_{ij}$. The unique coefficients were obtained via deforming the optimised structure, using a three-point formula, in the symmetrically required directions of both positive and negative amplitudes. These deformations correspond to tensile and compressive strains required to obtain the elastic response. The magnitude of each individual strain deformation is defined as 1%, ensuring the response is in the purely elastic region. For the visualization of the elastic tensors, and the calculation of the mechanical properties, the ELATE[56], EIAM[57], and Mathematica[58] software were used. Descriptions of the individual properties are given by Tan *et al.*[47]

ACKNOWLEDGEMENTS

A.F.M. thanks the Oxford Ashton Memorial scholarship for a DPhil studentship award. J.C.T. and A.F.M. are grateful for funding through the ERC Consolidator Grant (771575 (PROMOFS)) and the EPSRC Impact Acceleration Account Award (EP/R511742/1). We would like to acknowledge the use of the University of Oxford Advanced Research Computing (ARC) facility in carrying out this work (10.5281/zenodo.22558). Via our membership of the UK's HEC Materials Chemistry Consortium, which is funded by EPSRC (EP/R029431), this work used the ARCHER2 UK National Supercomputing Service (http://www.archer2.ac.uk). A.F.M. would like to thank Dr. Cyril Besnard and Prof. Alexander Korsunsky for their help with SEM imaging.

AUTHOR CONTRIBUTIONS

Experimental design, experimental execution, and data analysis: A.F.M; numerical simulations: L.D., A.F.M.; writing – original draft: A.F.M; writing – editing & reviewing: all authors; scientific input and supervision: B.C., J.C.T.

COMPETING INTERESTS

The authors declare no competing interests.